\documentclass[prd,onecolumn,aps,superscriptaddress,letterpaper,notitlepage]{revtex4-1}
\usepackage{amsmath,amssymb}
\usepackage{epsfig}
\usepackage{graphicx}
\usepackage{amsmath}
\usepackage{amsfonts}
\usepackage{epstopdf}
\usepackage{bbold}
\usepackage{color}
\usepackage{microtype,booktabs,multirow}
\usepackage{dsfont}  

\newcommand{\be}{\begin{equation}}
\newcommand{\ee}{\end{equation}}
\newcommand{\bea}{\begin{eqnarray}}
\newcommand{\eea}{\end{eqnarray}}

\newcommand{\beq}{\begin{equation}}
\newcommand{\eeq}{\end{equation}}

\newcommand{\ba}{\begin{array}}
\newcommand{\ea}{\end{array}}

\newcommand{\R}{{\mathds R}}
\newcommand{\C}{{\mathds C}}

\newcommand{\x}{\mathbf{x}}

\usepackage[colorlinks=true,backref=false, linktocpage=true,
citecolor=blue,urlcolor=blue,linkcolor=blue,pdfpagemode=UseOutlines]{hyperref}

\hypersetup{%
  bookmarksnumbered=true,
  pdftitle = {},
  pdfsubject = {},
  pdfauthor = {},
  pdfkeywords = {}
}

\let\Re\relax
\DeclareMathOperator{\Re}{Re}
\let\Im\relax
\DeclareMathOperator{\Im}{Im}

\def\beqs#1\eeqs{\beq\begin{split} #1 \end{split}\eeq}

\def\dd#1#2{\frac{d #1}{d #2}}
\def\pd#1#2{\frac{\partial #1}{\partial #2}}

\def\comment#1{}

\def\av#1{ \left\langle #1 \right\rangle }

\begin{document}

\title{Schwinger-Keldysh on the lattice: a faster algorithm and its application to field theory}

\author{Andrei Alexandru}
\email{aalexan@gwu.edu }
\affiliation{Department of Physics, The George Washington University,
Washington, DC 20052}
\affiliation{Department of Physics,
University of Maryland, College Park, MD 20742}
\author{G\"ok\c ce Ba\c sar}
\email{gbasar@uic.edu}
\affiliation{Department of Physics, University of Illinois, Chicago, IL 60607}
\author{Paulo F. Bedaque}
\email{bedaque@umd.edu}
\affiliation{Department of Physics,
University of Maryland, College Park, MD 20742}
\author{Gregory Ridgway}
\email{gregridgway@gmail.com}
\affiliation{Department of Physics,
University of Maryland, College Park, MD 20742}

\date{\today}

\begin{abstract}
A new algorithm is developed allowing the Monte Carlo study of a $1+1$ dimensional 
theory in real time.
The main algorithmic development is to avoid the explicit calculation of the Jacobian matrix 
and its determinant in the update process. This improvement has a wide applicability and 
reduces the cost of the update in thimble-inspired calculations from $\mathcal{O}(N^3)$ to 
less than $\mathcal{O}(N^2)$. 
As an additional feature, the algorithm leads to improved Monte Carlo proposals. 
We exemplify the use of the algorithm to the real time dynamics of a scalar $\phi^4$ theory 
with weak and strong couplings.
\end{abstract}

\pacs{}

\maketitle

\vskip0.2cm


\section{Introduction}
\label{sec:introduction}

Some of the most interesting -- and challenging -- problems in many-body physics are dynamical questions. They describe near equilibrium states through transport coefficients like diffusion constants, viscosities, conductivities, as well as phenomena away from equilibrium. The central objects of interest in these class of problems are the time dependent correlation functions of the form
\beq
\langle \mathcal{\hat O}(t) \mathcal{\hat O}(t')  \rangle ={\rm Tr} \left( \mathcal{\hat O}(t) \mathcal{\hat O}(t')  \hat\rho \right) \,.
\label{eq:rt_corr}
\eeq
where the operators have the usual time evolution $\mathcal{\hat O}(t) =e^{i \hat H t}\mathcal{\hat O}(0)e^{-i \hat H t}$ and $\hat\rho$ is the density matrix which reduces to the Boltzmann form, $e^{-\beta \hat H}/{\rm Tr} (e^{-\beta \hat H})$, in equilibrium. Problems in this category show up in almost every field of physics: cosmology, heavy ions collisions, and condensed matter physics to name a few.

Unfortunately the tools for tackling such dynamical problems from first principles are very limited. Even in weakly coupled systems, the study of long time dynamics (or low momentum properties) require complicated resummations of the perturbative expansion \cite{Braaten:1989mz,Jeon:1995zm}. Monte-Carlo techniques, the method of choice for non-perturbative problems, have a fundamental difficulty in dealing with real time (as opposed to imaginary time) dynamics, due to a particularly severe version of the ``sign problem". Many Monte Carlo-based approaches, including the one used in relativistic theories and that we use in this paper, are based on a path integral representation of the observable of interest. Such a path integral representation exists for real time observables in or out of equilibrium, and is based on the the Schwinger-Keldysh formalism \cite{Schwinger:1960qe,Keldysh:1964ud}. The problem in the Monte-Carlo evaluation of this path integral is that the integrand is a pure phase, as opposed to a fast decaying real function, and the importance sampling of the integrand, based on the interpretation of the integrand as a probability density, is not possible. As we will comment below, the sign problem for path integrals in the Schwinger-Keldysh formalism is, in a certain sense, the worst possible.

We are aware of two approaches that address this problem via Monte-Carlo techniques. The first one is to concentrate on near equilibrium and attempt to compute transport coefficients. They can be computed through the Kubo formula from the knowledge of certain equilibrium real time correlators \eqref{eq:rt_corr}. In principle, the correlators in imaginary time, $\langle \mathcal{\hat O}_E(\tau) \mathcal{\hat O}_E(\tau')  \rangle$ with $\mathcal{\hat O}_E(\tau) = e^{ \hat H \tau}\mathcal{\hat O}(0)e^{-\hat H \tau}$, contain the same information as the real time ones and can be computed with standard Monte Carlo techniques, frequently without a sign problem \cite{PhysRevD.35.2518,PhysRevD.47.653,Meyer:2007dy,Meyer:2007ic,Ding:2010ga}. In practice, however, exponentially good precision on  imaginary time is required to reconstruct it on real time. The second approach is to use Langevin methods (``stochastic quantization"\cite{:vn}). The drawback of the (complex) Langevin approach is that it does not always converge, or sometimes converges to an incorrect result~\footnote{For a recently proposed criterion for convergence see \cite{Nagata:2016vkn}}. In fact, the few attempts of applying the complex Langevin method to real time dynamics seem to suggest that it converges to the wrong result if the time separation $t-t'$ is more than the inverse temperature $\beta$ \cite{Berges:2005yt,Berges:2006xc,Mizutani:2008zz}. 

In the last few years a new approach to compute path integrals with sign problem has been developed \cite{Cristoforetti:2012su}. Although different versions vary in detail, they are all based in the deformation of the path integral from real values of the fields to a suitably chosen middle dimensional (i.e. with the same dimensions as the real field space) submanifold of the complexified field space. The equality of the integral over this new manifold to the integral over the original real space is guaranteed by a multidimensional version of Cauchy's theorem and by choosing the asymptotic properties of the manifold properly (the analogue of avoiding the ``arcs at infinity" familiar from complex analysis of functions of one complex variable). One choice of such a manifold is to deform the contour of integration from the real space to an appropriate combination of ``thimbles", the multidimensional analogues of the ``steepest descent" or ``constant phase" path from the theory of one complex variable. The sign problem is solved because, along thimbles, the phase of the integrand is constant. A difficulty with this choice of manifold is that it is, in general, nearly impossible to determine the particular combination of thimbles that is equivalent to the original region of integration. There are also the issues of how to sample disconnected thimbles and find them as their location is not known analytically.

Another choice of integration manifold was proposed in \cite{Alexandru:2015sua} and pursued by our group recently \cite{Alexandru:2016gsd,Alexandru:2016ejd,Alexandru:2017oyw}. In this proposal, the manifold of integration is obtained by taking the real fields as a starting value and evolving them according to the (anti)holomorphic gradient flow (the complex conjugate of the gradient of the action). This flow evolves a given real field configuration along a particular trajectory, determined by the gradient of the action, in the complexified field space. The end point of this evolution is determined by the ``flow time'', which is viewed as a free parameter. Therefore flowing the original real field space by some flow time creates an alternative, complex path integration manifold associated with the value of that flow time. In the limit of large flow times this manifold coincides with the precise combination of thimbles which is equivalent to the original integration domain. For finite flow times it provides a manifold i) that is equivalent to the original domain of integration, ii) on which the phase variation of the integrand is milder than on the real space and iii) that is connected, making the stochastic sampling easier to accomplish. 

This method was applied to the equilibrium real time dynamics of an anharmonic oscillator in \cite{Alexandru:2016gsd}. The correct result (which is known in this case through direct diagonalization of the hamiltonian) was obtained, even for time differences $t-t'$ of the order of $\approx 4 \beta$. However, the particular implementation of this method had two major shortcomings which prevented a similar computation in a field theory. 
Firstly,  an unreasonably large number of Monte-Carlo steps were required for thermalization and decorrelation. The reason for this was traced back to the fact that in this method, it is natural to make isotropic proposals in real space that are then ``flowed" to the manifold of integration where they are highly anisotropic. In \cite{Alexandru:2016gsd} some attempts were made to correct for this anisotropy by using a combination of gaussian approximation and trial-and-error adjustment for the proposals to make them more isotropic when flowed to the manifold of integration, but with limited success. It also required the computation and storage of a set of $N$, $N$-dimensional vectors (where $N$ is the number of degrees of freedom on the lattice), which was possible in the anharmonic oscillator problem but is prohibitive in a field theory with a large lattice.
Secondly, the need to compute the jacobian associated with the parametrization of the manifold by its real coordinates makes every step of the Monte-Carlo chain computationally expensive. Previously, this problem had been dealt with by using an estimator of the jacobian \cite{Alexandru:2016lsn,Alexandru:2016san,Alexandru:2016ejd} and reweighting the difference when making measurements. The estimator we developed is likely to be useful if the coupling is small and/or the manifold of integrations is nearly parallel to the real plane. This is not the case for the real time calculations so a new method that bypasses the need to compute the jacobian at every step of the Monte Carlo chain is necessary. 

The purpose of this paper is to present an algorithm without these two difficulties, and one that does not require the storage of the $N$, $N$-dimensional vectors. The main idea is the Grady algorithm \cite{Grady:1985fs,Creutz:1992xwa} which is used in lattice QCD in order to avoid the computation of fermion determinants. The effect of the jacobian is embedded in a bias of the proposals that are isotropic in the flowed manifold. We implement the Grady algorithm into the holomorphic gradient flow method to perform a real time calculation on a $\phi^4$ theory in $1+1$ dimensions. In this model and for weak or strong couplings, we are able to perform an even computationally cheaper calculation by approximating the holomorphic flow by its gaussian approximation in the calculation of the proposal (and reweighting the difference between them when making measurements).

In Section 2 we briefly review properties of the holomorphic flow and the new algorithm is described. In Section 3 the model and its representation in the Schwinger-Keldysh formalism is described. Results are described in Section 4 and a conclusion is presented in Section 5.


\section{ The Schwinger-Keldysh path integral and the model}

As discussed in the introduction, we are interested in expectation values of the form
\beq\label{eq:corr}
\langle \mathcal{\hat O}_1(x_1) \mathcal{\hat O}_2(x_2) \cdots \rangle
=
{\rm Tr}[\hat\rho(0) \mathcal{\hat O}_1(x_1) \mathcal{\hat O}_2(x_2) \cdots],
\eeq where $\mathcal{\hat O}_i(x)=e^{i\hat H t-i{\bf p}\cdot{\bf x}} \mathcal{\hat O}_i(t=0,{\bf x}={\bf 0}) e^{-i\hat H t+i{\bf p}\cdot{\bf x}}$ are operators in the Heisenberg representation 
and $\hat \rho(0) = e^{-\beta \hat H(0)}/{{\rm Tr} (e^{-\beta \hat H(0)})}$ is the density matrix representing the initial state of the system. When the time evolution is determined by a time independent hamiltonian $\hat H(t)=\hat H(0)=\hat H$, the system is in thermodynamic equilibrium which we will assume. The non-equilibrium case can also be studied within the formalism after a slight modification. We will briefly comment on this point later. Expectation values of this form can be obtained from the generating functional
\beq
Z[J_+, J_-, J_\beta] = {\rm Tr}[U(T-i\beta, T; J_\beta) U(T, T'; J_-) U(T', T; J_+)],
\label{eq:ZSK}
\eeq where $J_\pm, J_\beta$ are external classical currents coupled to the fields in the theory and $U(T,T';J)$ is the time evolution operator under the influence of the external current $J$ between times $T$ and $T'$. In order to compute correlators as in \eqref{eq:corr} we need $T < t_1, t_2, \cdots < T'$. The generating function has the path integral representation \cite{Schwinger:1960qe,Keldysh:1964ud}
\beq
Z[J_+, J_-, J_\beta] =\int D\phi_+ D\phi_- D\phi_\beta \, e^{i S_{SK}[\phi_+,\phi_-,\phi_\beta, J_+,J_-,J_\beta]} 
\eeq
where the action, $S_{SK}$, is defined as an integral of the Lagrangian along a time contour, ${\cal C}$, which lives in the complex plane:
\bea
S_{SK}[\phi_+,\phi_-,\phi_\beta, J_+,J_-,J_\beta]&=&\int_{\cal C}dt\,  \mathcal{L}[\phi_+,\phi_-,\phi_\beta, J_+,J_-,J_\beta]
\\
&=&  \int_T^{T'} dt \, \mathcal{L}[\phi_+, J_+] + \int_{T'}^{T'-i \beta/2} dt \, \mathcal{L}[\phi_\beta, J_\beta]
+   \int_{T'-i \beta/2}^{T-i \beta/2} dt  \, \mathcal{L}[\phi_-, J_-] 
\nonumber \\
&&+  \int_{T-i \beta/2}^{T-i \beta} dt\, \mathcal{L}[\phi_\beta, J_\beta]\,.\qquad
  \eea
%
with the boundary conditions $\phi_+(T')=\phi_\beta(T'), \phi_\beta(T'-i\beta/2)=\phi_-(T'-i\beta/2),\phi_-(T-i\beta/2)=\phi_\beta(T-i\beta/2), \phi_\beta(T-i\beta)= \phi_+(T)$ ($ \phi_\beta(T-i\beta)= -\phi_+(T)$ for fermionic fields).
\begin{figure}
\includegraphics[scale=0.5]{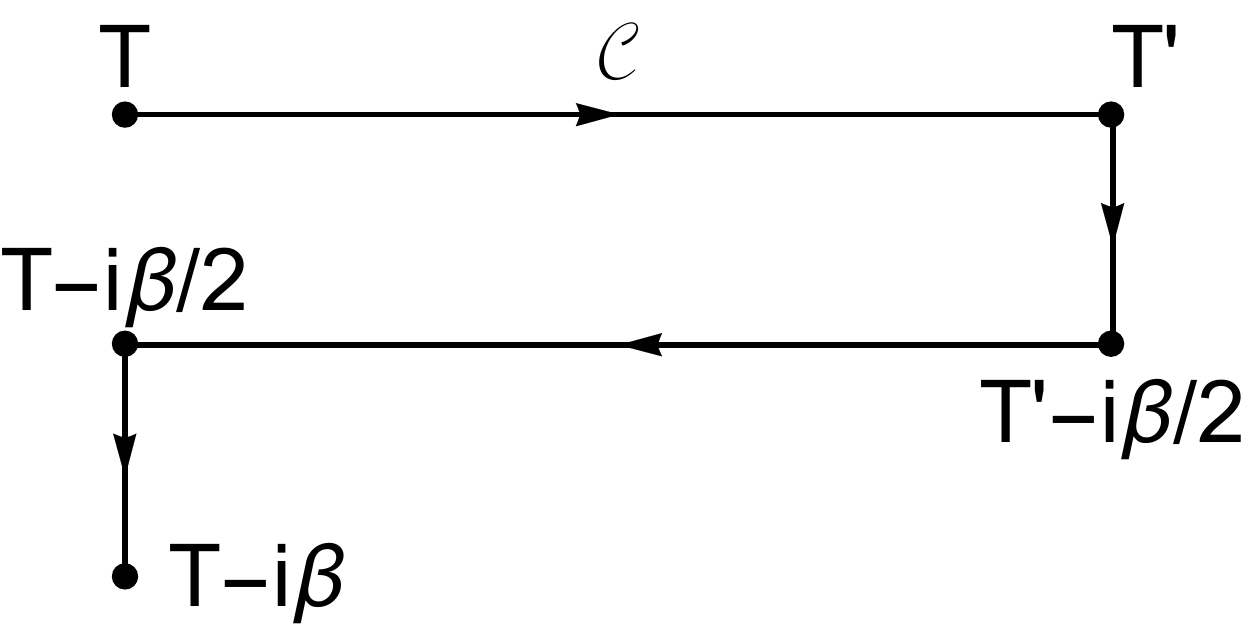}
\includegraphics[scale=0.4]{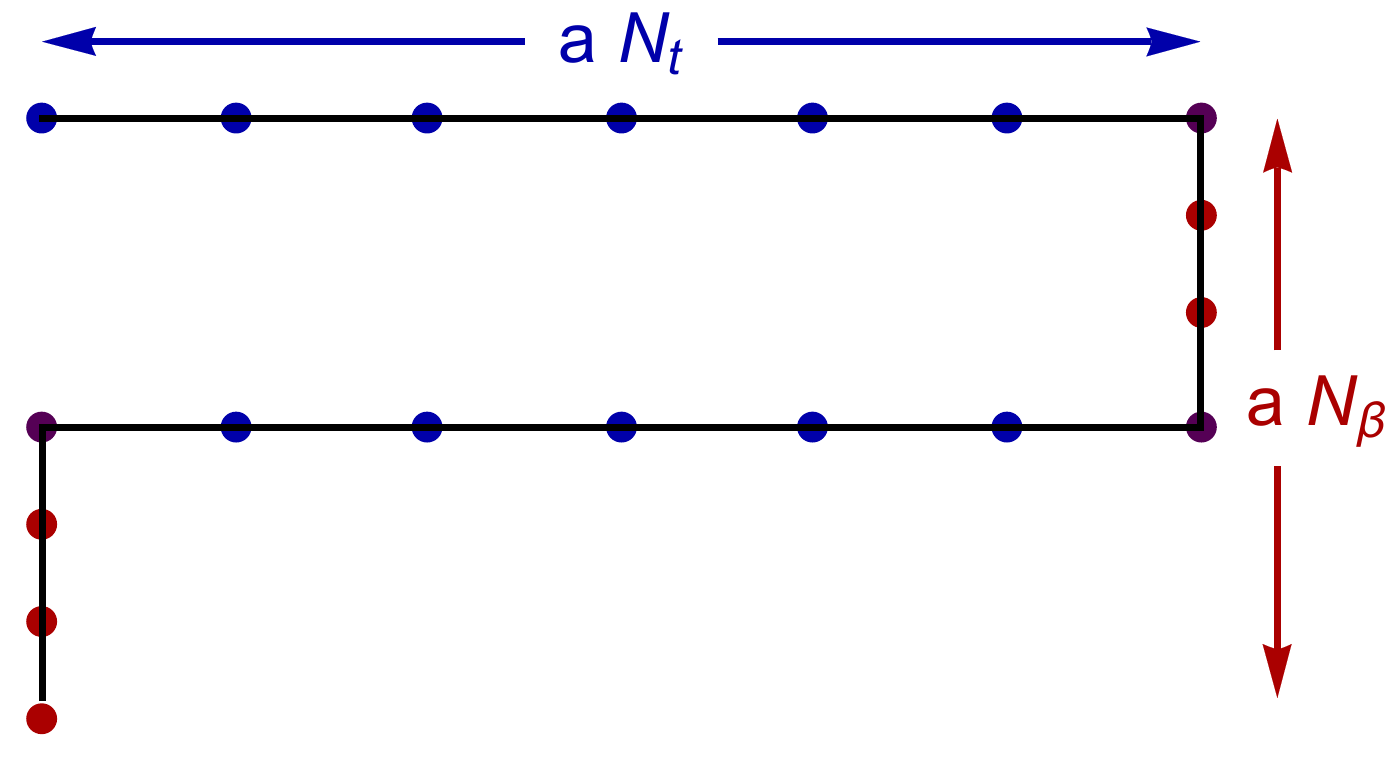}
\caption{The Schwinger Keldysh contour (left) and its discretization (right). }
\label{fig:sk}
\end{figure} 
The contour ${\cal C}$ is depicted in Fig.~\ref{fig:sk} (left). The upper and lower parts of the contour, parallel to the real axis, are associated with the forward and backward time evolution (i.e. the second and third terms in the trace in Eq.~\ref{eq:ZSK}). The parts that are along the imaginary axis are associated with the insertion of the density matrix (i.e. the first term in the trace in Eq.~\ref{eq:ZSK}). Note that we chose to split the density matrix into two parts that are inserted at times $T$ and $T'$. This choice assumes the existence of equilibrium where the Hamiltonian is time independent and the density matrix commutes with time evolution operator. In order to study an out-of-equlibrium system this contour has to be modified such that the density matrix is inserted at time $T'$ as a whole. Even though our construction can be generalized, we will not discuss this case in this paper. 
 
 Given the path integral representation, the various correlators can be computed  by differentiating $Z[J_+, J_-, J_\beta] $ with respect to external sources. The time ordering is such that the operators that are inserted in the lower branch of ${\cal C}$ always have a larger time compared to those that are inserted at the upper branch. A two point correlator with both operators are in the upper (lower) branch is time ordered (anti time ordered). For instance 
 \bea
 \label{eq:sk_2pt}
\langle  T \phi(t_1,\x_1)  \phi(t_2,\x_2)   \rangle = \frac{\delta^2 Z[J_+, J_-, J_\beta] }{\delta J_+(t_1,\x_1)  \delta J_+(t_2,\x_2)}
\,.
\eea

In the paper, we present a Monte Carlo method to compute time dependent correlation functions as in \eqref{eq:sk_2pt}  for the $1+1$ dimensional $\phi^4$ theory with the potential $V[\phi]={1\over 2}m^2\phi^2+{\lambda \over 4!} \phi^4$. The lattice action corresponding to the Schwinger-Keldysh path integral is given by  
\beqs
\label{eq:sk_discrete}
S[\phi] \equiv -i S_{SK,lattice} 
= \sum_{t,n} a_t a  & \left[
 \frac{(\phi_{t+1,n}-\phi_{t,n})^2}{2 a_t^2} 
+ \frac12 \left( \frac{(\phi_{t+1,n+1}-\phi_{t+1,n})^2}{2 a^2} 
+  \frac{(\phi_{t,n+1}-\phi_{t,n})^2}{2 a^2} 
\right) \right.
\\ 
&
+\left. \frac12 m^2 \frac{ \phi_{t,n}^2 + \phi_{t+1,n}^2}{2}
 + \frac{\lambda}{4!} \frac{\phi_{t+1,n}^4 +  \phi_{t,n}^4 }{2}
\right] \,,
\eeqs 
where $t$ and $n$ indexes the lattice along the time and spatial directions,  $a$ is the spatial lattice spacing and $a_t$ is the time lattice spacing:
\beqs
a_t =& ia, \quad \text{for} \quad 0\leq t< N_t, \\
a_t =& a, \quad \text{for} \quad N_t\leq t< N_t+N_\beta/2, \\
a_t =& -ia,\quad \text{for} \quad  N_t+N_\beta/2\leq t< 2N_t+N_\beta/2,  \\
a_t =& a, \quad \text{for} \quad 2N_t+N_\beta/2\leq t< 2N_t+N_\beta,
\eeqs 
and $N_t, N_\beta$ are the number of lattice points on the real and imaginary axis, respectively. This is shown in Fig.~\ref{fig:sk}~(right). We embedded a factor of $-i$ in the definition of the action in Eq.~\ref{eq:sk_discrete} so that the measure in the path integral is $e^{-S}$.  The change in $a_t$ with $t$ determines the  contour in the complex time plane defining the Schwinger-Keldysh action. The fields $\phi_+, \phi_-$ and $\phi_\beta$ correspond, in our discretized action, to $\phi_{t,n}$ for $t$ in the ranges $0\leq t< N_t$, $ N_t+N_\beta/2\leq t< 2N_t+N_\beta/2$, and $  N_t\leq t< N_t+N_\beta/2 $ and $2N_t+N_\beta/2\leq t< 2N_t+N_\beta/2$ respectively. The correlators that we are interested in computing are now given by the discretized path integral 
\bea
\langle \phi_{t_1,n_1} \phi_{t_2,n_2}   \rangle = {\int \big(\prod_{t,n}d\phi_{t,n} \big) e^{-S[\phi]} \phi_{t_1,n_1} \phi_{t_2,n_2}\over \int \big(\prod_{t,n}d\phi_{t,n} \big) e^{-S[\phi]}} \,.
\eea

Along the two branches of the real axis the measure, $e^{-S[\phi]}$, of the path integral is a pure phase, and lacks exponential damping as the value of $\phi_\pm(t,\x)$ is varied, in contrast to the Euclidean branch. Thus, any attempt at reweighting the phase of the integrand is doomed to fail as the average phase  vanishes. In this sense, the sign problem arising in real time is the worst possible. In the following two sections, we will present a new Monte-Carlo algorithm to alleviate this sign problem by using the holomorphic gradient flow. 

In order to validate our results at small coupling and to ascertain that our results at larger coupling indeed cannot be reproduced by perturbation theory, we compare against perturbative results.
The propagator in the Schwinger-Keldysh contour is given up to order 
${\cal O}(\lambda^2)$ by
\beqs
\langle \phi_i \phi_j\rangle 
&\approx
 \langle \phi_i \phi_j\rangle _0
+\lambda \left( -\frac{1}{4!}\right)\sum_k \tilde a_k \langle \phi_i \phi_j \phi_k^4\rangle _0^c
+\lambda^2 \frac12 \left( -\frac{1}{4!}\right)^2\sum_{k,l} \tilde a_k \tilde a_l \langle \phi_i \phi_j \phi_k^4 \phi_l^4\rangle _0^c \\
&=
(H_0)^{-1}_{ij}
-\frac{\lambda}{2} \sum_k \tilde a_k 
(H_0)_{ik}^{-1}(H_0)_{kk}^{-1}(H_0)_{kj}^{-1}
+\frac{\lambda^2}{4} \sum_{k,l} \tilde a_k \tilde a_l   \left[ 
  (H_0)_{il}^{-1}((H_0)_{lk}^{-1})^2(H_0)_{kk}^{-1}(H_0)_{kj}^{-1}\right.\\
&\phantom{+\frac{\lambda^2}{4} \sum_{k,l} \tilde a_k \tilde a_l   (H_0)_{il}^{-1}((H_0)_{lk}^{-1})^2}
 \left.
+ (H_0)_{il}^{-1}(H_0)_{lk}^{-1}(H_0)_{ll}^{-1}(H_0)_{kk}^{-1}(H_0)_{kj}^{-1}
+\frac{2}{3} (H_0)_{il}^{-1}((H_0)_{lk}^{-1})^3(H_0)_{kj}^{-1}
\right],
\eeqs 
where $i,j,k,l$ are combined time and space indices, 
$\tilde a_k = a(a_{t(k)-1} + a_{t(k)})/2$, $H_0$ is the Hessian of the 
Schwinger-Keldysh action at $\phi=0$ and $\langle \cdot\rangle_0^c$ 
denotes the connected part of the correlator.


\section{Holomorphic gradient flow}

In order to solve the sign problem, that is to reduce the phase fluctuations for
the path integral to a level where we can easily reweight it, we will
 deform the integration manifold of our path integral. The first step
of the process is to promote all real variables (values of the field
at each space-time point) to complex ones. Using a generalized version of
 Cauchy's theorem, we can show that we can deform the integration
manifold in the complex space without changing the value of the integral,
as long as we do not cross any singularities of the integrand and we preserve 
the asymptotic behavior of the field. There is a rather large latitude
 in choosing the deformation but in this work we will use a 
deformation induced by the {\em holomorphic gradient flow}. In this section 
we will review briefly the relevant details.

The holomorphic gradient flow is defined, for a system with $N$ real degrees
of freedom, through the set of differential equations
\beq
\dd{z^i(\tau)}{\tau}  = \overline{\pd{S(z(\tau))}{z^i}}\quad\text{with}
\quad z(0)  = x\in \R^N \,.
\label{eq:hgf}
\eeq
Integrating the flow equation above for a fixed amount of ``time'' $T_\text{flow}$ 
defines a map $x\to f(x)$ where $f(x)\equiv z(T_\text{flow})$. 
The image under this map of the original integration domain $\R^N$ 
is our new {\em integration manifold} ${\cal M}=f(\R^N)$. Note that this
manifold depends on $T_\text{flow}$ and as the flow time is increased
the sign fluctuations become milder~\cite{Alexandru:2015xva}. When the action
is real the flow keeps the points in the real subspace, but for complex actions
the image manifold ${\cal M}$ will be different from $\R^N$. Since we
use the points in $\R^N$ to keep track of the points in the integration
manifold ${\cal M}$, we will refer to $\R^N$ as the {\em parametrization
manifold}.

\def\Tf{{\rm T}\!f}
The map $f$ naturally induces a linear map, denoted by $\Tf$, 
between the tangent space at
point $x$ and the tangent space at point $x'=f(x)$. A vector $v$
tangent at $x$ is mapped to $\Tf(v)=v'$ tangent at $x'$ with
$v' = \omega(T_\text{flow})$ where $\omega$ satisfies the
differential equation
\beq
\dd{\omega^i(\tau)}{\tau} = \overline{H_{ij}(z(\tau)) \omega^j(\tau)} \quad\text{with}
\quad \omega(0)=v\quad\text{and}
\quad H_{ij}\equiv \frac{\partial^2S}{\partial z^i \partial z^j} \,.
\label{eq:vhgf}
\eeq
This equation is derived by considering the flow of a point infinitesimally 
displaced in direction $v$ away from $x$.
Note that in the equation above the Hessian $H$ is to be evaluated along the trajectory $z(\tau)$
that takes $x$ to $x'$, so this equation needs to be solved together with Eq.~\ref{eq:hgf} in order to ``transport" a vector.

Since the vector flow is a linear map, it can be represented by a matrix.
Taking in $\R^N$ the canonical basis $\{e_j\}_{j=1,\ldots,N}$ with $e_j^i = \delta_{ij}$ the
vectors $\Tf(e_j)$ will form a basis for the tangent space of $\mathcal{M}$ at $x'$. The
matrix $J$ which has these vectors as columns, that is
\beq
J_{ij} \equiv \Tf(e_j)^i \,,
\label{eq:J}
\eeq
can be used to map $v=v^i e_i$ into $v'= v^i \Tf(e_i)$ so that the components
of $v'$ are $v'^i = J_{ij} v^j$. Note that while the components of $v$ have
to be real, the components of $v'$ are in general complex, since the tangent
space at $x'$ is usually not parallel to $\R^N$. We stress that $v'^i$ are
the components of $v'$ in the canonical basis $e_i$ in $\C^N$, not the 
components of $v'$ in the basis $\Tf(e_i)$. If we decompose $v'$ in the
basis $\Tf(e_i)$ the components are real, as they should be, and they are
in fact $v^i$.
The map $\Tf$ and the matrix $J$ depend on the starting point $x\in\R^n$,
but to simplify the notation we will make this explicit only when required.
 
%

An important property of the tangent map $\Tf$ is that it preserves the 
imaginary part of the dot product of two vectors since the derivative of the
product along the flow is real:
\beq
\dd{}\tau \av{w,v}_\C= \frac{d}{d\tau}(\overline{w}^iv^i) 
= \overline{\frac{dw}{d\tau}}^iv^i + \overline{w^i}\frac{dv}{d\tau}^i 
= H_{ij}w^jv^i + \overline{w^iH_{ij}v^j} 
=  2\text{Re}[w^iH_{ij}v^j] \,.
\eeq
We denoted the $\C^N$ scalar product with $\av{w,v}_\C$ to distinguish
it from the real scalar product 
$\av{w,v}_\R=\Re w^i \Re v^i+\Im w^i\Im v^i=\Re\av{w, v}_\C$.
The invariance of the imaginary part implies that for any $i$ and $j$
\beq
\Im\av{\Tf(e_i),\Tf(e_j)}_\C = \Im\av{e_i,e_j}_\C=0\quad\text{which can be
restated as}\quad \Im\overline{J_{ki}}J_{kj}=\Im(J^\dagger J)_{ij}=0 \,.
\eeq
Thus the matrix $J^{\dagger}(x)J(x)$ is real.
Another important property is that while $\{\Tf(e_j)\}_{j=1,\ldots,N}$ span
the tangent space at $x'$ (viewed as a real vector space), the vectors 
$\{i\Tf(e_j)\}_{j=1,\ldots,N}$ span the orthogonal vector space. This
can be seen by considering the real scalar product between any two vectors in
these sets:
\beq
\av{\Tf(e_j),i\Tf(e_k)}_\R = \Re\av{\Tf(e_j),i\Tf(e_k)}_\C=
-\Im\av{\Tf(e_j),\Tf(e_k)}_\C = 0 \,.
\eeq 
This means that any vector $v'$ at $x'$ can be decomposed as 
\beq
v'=v'_\parallel+ v'_\perp\quad\text{with}\quad
v'_\parallel=v_\parallel^j \Tf(e_j)=\Tf(v_\parallel^j e_j)\quad\text{and}\quad
v'_\perp=v_\perp^j i\Tf(e_j)=i\Tf(v_\perp^j e_j)\,,
\label{eq:decomp}
\eeq 
where $v_{\parallel,\perp}^j$ coefficients are real. Defining 
$v_{\parallel,\perp}\equiv v_{\parallel,\perp}^j e_j$, we have 
$v'_\parallel=J v_\parallel$ and $v'_\perp = i J v_\perp$.


The partition function is evaluated by integrating over the manifold ${\cal M}$:
\beq
Z=\int_{\cal M}\! {\rm d}x'\, e^{-S(x')} = \int_{\R^N}\! {\rm d}x\, \det J(x)\, e^{-S(f(x))} = \int_{\R^N}\! {\rm d}x\, |\det\,J(x)| e^{-\Re S(f(x))} \Phi(x)\,,
\eeq
where $\Phi(x)\equiv\exp[-i \Im S(f(x))+ i\arg \det\,J(x)]$ is a pure phase. 
Note that the measure $dx'$ is the Cauchy measure on the manifold ${\cal M}$
embedded in $\C^N$ and that the change of variables from the integration
manifold to the parametrization manifold $\R^N$ gives rises to the Jacobian
$\det\,J(x)$, which is the determinant of the matrix $J$ defined in Eq.~\ref{eq:J}.
To evaluate observable averages with respect to $Z$, we will sample 
configurations according to the positive 
weight $P(x)$ and then evaluate observables by reweighting, that is, we compute averages from the formula
\beq
\av{{\cal O}(x)} = \frac{\av{{\cal O}(x) \Phi(x)}_P}{\av{\Phi(x)}_P}
\quad\text{with}\quad
P(x)= \left|\det\,J(x)\right| e^{-\Re S(f(x))} \,.
\label{eq:P}
\eeq
The averages $\av{\cdot}_P$ are taken with respect to the probability
weight $P(x)$. We discuss how to sample this measure in the next section.


\section{Algorithm}
\label{sec:algo}

In this section we will present a method of sampling configurations according
to the probability weight $P(x)$ defined in Eq.~\ref{eq:P}. 
The method described here is based on the Metropolis algorithm, where
new configurations are proposed and an accept-reject step is used
to ensure detailed balance. The algorithm we used previously ~\cite{Alexandru:2015xva} suffered from two main flaws. The first was the substantial  cost of computing the jacobian $J(x)$ at every step of the Markov chain. The second was that the proposals, straightforwardly, were chosen to be isotropic in the real variables $x$. As those variables parametrize the actual manifold of integration $\mathcal{M}$ through the very non-linear map $x'=f(x)$, the resulting proposals were very anisotropic in $\mathcal{M}$. This distortion effect was partially, but not completely,  compensated in ~\cite{Alexandru:2015xva}. The result was that very small proposal steps had to be chosen in order to achieve reasonable acceptance rates and made the algorithm perform poorly. The algorithm we discuss here improves on the one used in ~\cite{Alexandru:2015xva} in both respects.


To better understand the advantages of this algorithm, let us first review
 the computationally intensive steps involved. The cost estimates below
will concern typical bosonic systems, as the one considered in this
paper. The most expensive part of the computation of the action of the
flowed configuration $S(x')=S(f(x))$, is the
calculation of the flowed configurations $x'$, which requires the integration
of Eq.~\ref{eq:hgf}. For this, we use an adaptive Runge-Kutta integrator~\cite{Cash:1990:VOR:79505.79507}. 
In terms of scaling with the size of the system this is a 
${\cal O}(N)$ calculation, if we assume that $T_\text{flow}$ remains fixed
as we increase the system size. To compute $J(x)$ we need to integrate
Eq.~\ref{eq:vhgf} for each vector in the basis. The Hessian is usually 
a sparse matrix, so each integration step can be implemented
with complexity ${\cal O}(N)$, and the cost of computing $\Tf(v)$ for
some vector $v$ is ${\cal O}(N)$. Overall, the cost of computing $J(x)$ is
then ${\cal O}(N^2)$ and its determinant has a cost of order ${\cal O}(N^3)$. For large
systems the cost of the later steps quickly becomes dominant.
For some systems estimators of $\left|\det\, J(x)\right|$ can be employed for sampling
and the difference can be reweighted~\cite{Alexandru:2016lsn}. However, these
estimators do not work well for the system considered in this paper.



\def\Pr{{\rm Pr}}
\def\Pacc{{{\rm P}_\text{acc}}}

The second problem is to find proposals that move efficiently through the integration
manifold ${\cal M}$. Proposals that are distributed isotropically in the 
parametrization space are in general mapped to a highly skewed distribution in 
the integration space, leading to an inefficient sampling of the 
manifold~\cite{Alexandru:2015xva}. The skewed distribution appears because
the map $\Tf$ scales very differently vectors that point in different directions,
that is, the eigenvectors of $J(x)$ have eigenvalues of very different
magnitudes. When the matrix $J(x)$ is relatively constant over the sampled
region in the parametrization space, we can bias the proposals in the parametrization
space such that their distribution in the integration manifold is relatively
isotropic. To be specific, denote the current configuration by $x_n$
and the proposed one by $x_{n+1}$. When the proposal is selected with 
probability $\Pr(x_{n}\to x_{n+1})\propto\exp(-\Delta^T M \Delta)$, where 
$\Delta\equiv x_{n+1}-x_n$ and $M$ is a fixed real positive-definite matrix, 
the probability $\Pr$ is symmetric in $x_n$ and $x_{n+1}$,
so the acceptance probability required for detailed balance is the usual
$\Pacc=\min\{1, P(x_{n+1})/P(x_n)\}$. When the parametrization manifold
is tangent to a critical point $x_\text{cr}$, a good choice for matrix
$M$ is $J(x_\text{cr})^\dagger J(x_\text{cr})$, since this bias can
be effectively constructed using the ``eigenvectors'' and 
``eigenvalues'' of the Hessian evaluated at the critical 
point~\cite{Alexandru:2015xva,Alexandru:2016san}. When flowing from the original
integration manifold $\R^N$, as we will do in the present paper, a possible choice for the matrix $M$
would be the quadratic approximation to the real part of the action
$\Re S(f(x))\approx -x^T M x$~\cite{Alexandru:2016gsd}. 
The problems with these methods are that i) the accept-reject step requires the 
calculation of $J(x)$ and its determinant, and ii) they are only effective
when the matrix $J$ does not fluctuate too much over the sampled configurations.
Ideally, we would like to make proposals that are isotropic around $x_n'\in{\cal M}$,
that is $\Pr(x_{n}\to x_{n+1})\propto\exp(-\Delta^T J^\dagger(x_n)J(x_n) \Delta)$.
These proposals are isotropic in the tangent space at $x_n'$ because 
$\eta'_\parallel\equiv J(x_n)\Delta$ is a random vector in this space distributed 
with probability 
$P(\eta_\parallel')\propto\exp(-\eta_\parallel'^\dagger \eta_\parallel')$.
These proposals are not symmetric under the exchange $x_n \leftrightarrow x_{n+1}$, since the matrices
$J_n\equiv J(x_n)$ and $J_{n+1}\equiv J(x_{n+1})$ are different. 
To account for this asymmetry, 
the acceptance probability needs to be modified to
$\Pacc=\min\{1, P(x_{n+1}) \left|\det J_{n+1}\right|/
P(x_n) \left|\det J_{n}\right|\}$.
This is still expensive, since the determinants appearing in $P(x)$ do not
cancel in the acceptance ratio and we are still required to compute 
$\left|\det J\right|$. However, this suggests a way to arrange the proposals
to cancel the determinants: if we could make proposals using the probability
$\Pr(x_{n}\to x_{n+1})\propto\exp[-\Delta^T (J^\dagger J)_{n+1} \Delta]$
computing the acceptance would require just the action
difference $\Delta S=\Re[S(x_{n+1}')-S(x_n')]$. This would require solving 
an implicit equation to determine the new configurations $x_{n+1}$. 
This is not the approach we will follow here. Instead, we follow
a method based on the Grady algorithm~\cite{Grady:1985fs,Creutz:1992xwa}
which can be designed to both be isotropic around $x_n'$ and also
to avoid explicit computation of $J$; we make proposals using
\beq
\Pr(x_n\to x_{n+1}) = \sqrt{\frac{\det (J^\dagger J)_{n}}{\pi^N \delta^2}} e^{-\Delta^T (J^\dagger J)_n\Delta/\delta^2} \,,
\eeq
where $\delta$ is a parameter used to adjust the step size so that the
acceptance rate is reasonable. Note that this steps relies on the fact that the
matrix $J^\dagger J$ is real, a property discussed earlier in the paper.
The equation above involves $\left|\det\, J\right|$, but this will appear
in our calculation only implicitly, since our algorithm will require only
that we generate vectors $\eta_\parallel' = J_n\Delta$ that are normally distributed; we
only displayed the full probability function above to help us prove the
detailed balance below.
The acceptance rate is computed by first generating
an auxiliary complex vector $\xi$ with probability
\beq
P(\xi) = \frac{\det (J^\dagger J)_{n+1}}{\pi^N} e^{-\xi^\dagger(J^\dagger J)_{n+1}\xi} \,,
\label{eq:pxi}
\eeq
and then the acceptance probability is computed using
\beq
\Pacc = \min\{1, e^{-\Delta S+\xi^\dagger[(J^\dagger J)_{n+1}
-(J^\dagger J)_{n}]\xi+\Delta^T[(J^\dagger J)_{n}
-(J^\dagger J)_{n+1}]\Delta/\delta^2} \} \,.
\label{eq:pacc}
\eeq
The total transition rate from $x_n$ to $x_{n+1}$ is then
\beq
T(x_n\to x_{n+1}) = \Pr(x_n\to x_{n+1}) \int {\rm d}\xi{\rm d}\xi^\dagger\,
 P(\xi) \Pacc \,.
\eeq
To prove the correctness of the method we follow a strategy outlined 
by Creutz~\cite{Creutz:1992xwa}: we show that the product 
$P(x_n)T(x_n\to x_{n+1})$ is symmetric in $x_n$ and $x_{n+1}$ which 
implies that the detailed balance is satisfied. We have
\beqs
P(x_n)T&(x_n\to x_{n+1}) = e^{-\Re S(x_n')} \left|\det\,J_n \right| 
\sqrt{\frac{\det (J^\dagger J)_{n}}{\pi^N \delta^2}} e^{-\Delta^T (J^\dagger J)_n\Delta/\delta^2} \\
&\times\int {\rm d}\xi{\rm d}\xi^\dagger\,
\frac{\det (J^\dagger J)_{n+1}}{\pi^N} e^{-\xi^\dagger(J^\dagger J)_{n+1}\xi}
 \min\{1, e^{-\Delta S+\xi^\dagger[(J^\dagger J)_{n+1}
-(J^\dagger J)_{n}]\xi+\Delta^T[(J^\dagger J)_{n}
-(J^\dagger J)_{n+1}]\Delta/\delta^2} \} \\
=&\frac{\det(J^\dagger J)_n \det(J^\dagger J)_{n+1}}{\delta \pi^{3N/2}} \\
&\times\int {\rm d}\xi{\rm d}\xi^\dagger\,
\min\{
e^{-\Re S(x_n')-\Delta^T(J^\dagger J)_n \Delta/\delta^2-\xi^\dagger
(J^\dagger J)_{n+1}\xi}, 
e^{-\Re S(x_{n+1}')-\Delta^T(J^\dagger J)_{n+1} \Delta/\delta^2-\xi^\dagger
(J^\dagger J)_{n}\xi} \}
\,.
\eeqs
This proves the correctness of our method. To derive the relation above we used 
the fact that $J^\dagger J$ is real and positive-definite, so that 
$\sqrt{\det J^\dagger J}=\left|\det J\right|$.

 Our algorithm samples configurations $x\in\R^N$ distributed with the desired probability $P(x)$
given in Eq.~\ref{eq:P}. The configurations are updated using the following steps:
\begin{enumerate}
\item Generate a complex vector $\eta'\in\C^N$ using the distribution 
$P(\eta')\propto\exp(-\eta'^\dagger\eta'/\delta^2)$.
\item Compute $\Delta=\Re J_n^{-1}\eta'$ and $\eta'_\parallel=J_n\Delta$ which is  the tangent component of $\eta'$ at $x_n'$. 
\item Propose a new configuration $x_{n+1}=x_n+\Delta$.
\item Generate a complex vector $\zeta'\in\C^N$ using the probability distribution 
$P(\zeta')\propto\exp(-\zeta'^\dagger \zeta')$.
\item Compute $\xi=J_{n+1}^{-1}\zeta'$ and use it to determine whether to 
accept the new configuration using $\Pacc$ in Eq.~\ref{eq:pacc}.
\end{enumerate}
In step 1 and 4, $\eta'$ and $\zeta'$ can be generated by simply drawing each component from an  appropriate normal distribution. 
In step 2, in order to show that $\eta'_\parallel$ is tangent to $\mathcal{M}$ at $x'$ 
 we used the fact that we can decompose $\eta'$ into components parallel and tangent: 
$\eta'=\eta_\parallel'+\eta_\perp'=J(\eta_\parallel+
i\eta_\perp)$, where $\eta_{\parallel,\perp}\in\R^N$, and thus $\Delta=\eta_\parallel=\Re J_n^{-1}\eta'$, as it  follows from the discussion surrounding Eq.~\ref{eq:decomp}.

The final ingredient necessary to implement steps 2 and 5 is a method to compute $J^{-1}v'$ for an arbitrary complex
vector $v'\in\C^N$ \textit{without} first evaluating the matrix $J$. We note first
that, for a real vector $v$, we can compute $Jv$ simply by integrating 
Eq.~\ref{eq:vhgf} to get $\Tf(v)$, a calculation of complexity ${\cal O}(N)$. However,
this does not work when the vector $v$ is complex, because the flow is 
non-linear when the vector has imaginary components, that 
is $\Tf(v+i w)\neq\Tf(v)+i\Tf(w)$. 
For a complex vector the solution is to evolve the real and imaginary 
components separately, that it $Jv = \Tf(\Re v)+i\Tf(\Im v)$. This requires
two separate integrations of Eq.~\ref{eq:vhgf}, but the complexity 
remains ${\cal O}(N)$. Armed with this routine, we can then compute
$J^{-1}v'$ for any vector by using an iterative method, for example 
GMRES~\cite{gmres} or BiCGstab~\cite{vorst:631}. These algorithms allows
us to compute  $J^{-1}v'$ through successive computations of $J v$
 without explicitly inverting $J$. We note that when the vector
$v'$ is tangent at $x'$, its inverse $J^{-1}v'\in\R^N$ can be computed
by integrating Eq.~\ref{eq:vhgf} backwards. While we did not use this 
property for this study, it is conceivable that this may be employed to optimize an inversion
algorithm for general vectors.

The algorithm described in this section can be used to sample $P(x)$
without computing the matrix $J(x)$ and its determinant. To complete
the calculation we need to evaluate the phase $\varphi(x)$. While
there are methods that can estimate this phase accurately without 
computing $\det J$~\cite{Cristoforetti:2014gsa}, in
this paper we evaluate it directly by computing $\det J$ on the 
decorrelated configurations, which is a small subset of the one
generated by the sampling algorithm.

Later on, in order to gauge the performance of our algorithm, we will compare it with
a variant of our algorithm. For that purpose we improve the way proposals are made in relation to the method used in \cite{Alexandru:2016gsd}. The proposals are made according to the distribution
$\Pr(x_{n}\to x_{n+1})\propto\exp(-\Delta^T J^\dagger(0)J(0) \Delta/\delta^2)$ 
and accepted with probability $\Pacc=\min\{1, \exp(-\Delta S)\}$. This in effect samples
the configurations not with the desired probability but with probability 
$P_0(x)\propto\left|\det J(0)\right|\exp(-\Re S(f(x)))$ so, when computing he observables the difference between the desired probability and $P_0(x)$ has to be reweighted by using
\beq
\av{{\cal O}} = \frac{\av{{\cal O}(x)\tilde\Phi(x)}_{P_0}}
{\av{\tilde\Phi(x)}_{P_0}}
\quad\text{with}\quad
\tilde\Phi(x) = \frac{\det J(x)}{\left|\det J(0)\right|}e^{-i\Im S(f(x))} \,,
\eeq
where the average $\av{\cdot}_{P_0}$ is taken with respect to the 
weight $P_0(x)$. For the model discussed in this paper, at weak coupling the 
matrix $J(x)$ will be close to $J(0)$ and the method should work well. As
we increase the coupling the reweighting factor $\tilde\Phi$ oscillates
quickly and the statistical power of the ensemble generated by $P_0$ will 
decrease making the process inefficient. We note that for this method
we need  to compute $J(0)^{-1}v$ for a large numbers of vectors. 
For sufficiently small systems, as the ones we consider in this paper,
this can be done efficiently by precomputing $J(0)^{-1}$,
allowing us to generate a large number of configurations very cheaply.



\section{Results}
\label{sec:results}

The data presented in this section refers to simulations that are run on
lattices with $N_t=8$, $N_\beta=2$, and $N_x=8$. The total
number of degrees of freedom for this system is 
$(2\times N_t+N_\beta)\times N_x=144$. We set $m=1$
and the lattice spacing $a=0.2$. We run simulations for three
different values of $\lambda$: $0.1$, $0.5$, and $1.0$. 
For each $\lambda$ we did short runs to determine the average 
sign $\Re\av{\Phi}_P$ and increased $T_\text{flow}$ until 
it reached a value of around $0.2$. The
parameters for these simulations, including the flow time
$T_\text{flow}$, the average sign, and the number of updates
for both our current algorithm and the old one are collected
in Table~\ref{tab:simulations}.

\begin{table}[b]
\centering
\begin{tabular*}{0.8\textwidth}{@{\extracolsep{\stretch{1}}}ccccccccccc@{}}
\toprule
\multirow{2}{*}{$\lambda$} &\phantom{a}& \multirow{2}{*}{$T_\text{flow}$} & 
\multirow{2}{*}{$\Re\av{\Phi}_P$} & \multicolumn{4}{c}{$J_0$} &\multicolumn{3}{c}{Grady}\\
\cmidrule{5-8} \cmidrule{9-11}
&&&& ${\rm updates}/10^6$ & 
$\delta$& acceptance& stat power &$\text{updates}/10^6$ & $\delta$ & acceptance\\
\midrule
0.1    && 1.8   & 0.278(2)               & 100  & 0.11 & 53\%       &    0.99        & 3.2                         & 0.1 & 57\%\\
0.5    && 1.6   & 0.193(2)              & 50     & 0.115 & 50\%      &   0.90        & 2.4                         & 0.1 & 57\%\\
1.0     && 1.6   & 0.189(2)              & 50      & 0.1 & 53\%       &  0.68       & 2.4                         & 0.1 & 55\%\\ \bottomrule
\end{tabular*}
\caption{Simulation parameters for the ensembles used in this study. The last
two sets of columns indicate the number of updates required to generate the 
ensembles used in this study, the step-size $\delta$, and the acceptance rate. 
Grady indicates the algorithm proposed in this
paper, and $J_0$ indicates the variant of the old algorithm described at
the end of Section~\ref{sec:algo}.}
\label{tab:simulations}
\end{table}

\begin{figure}[t]
\includegraphics[width=0.9\textwidth]{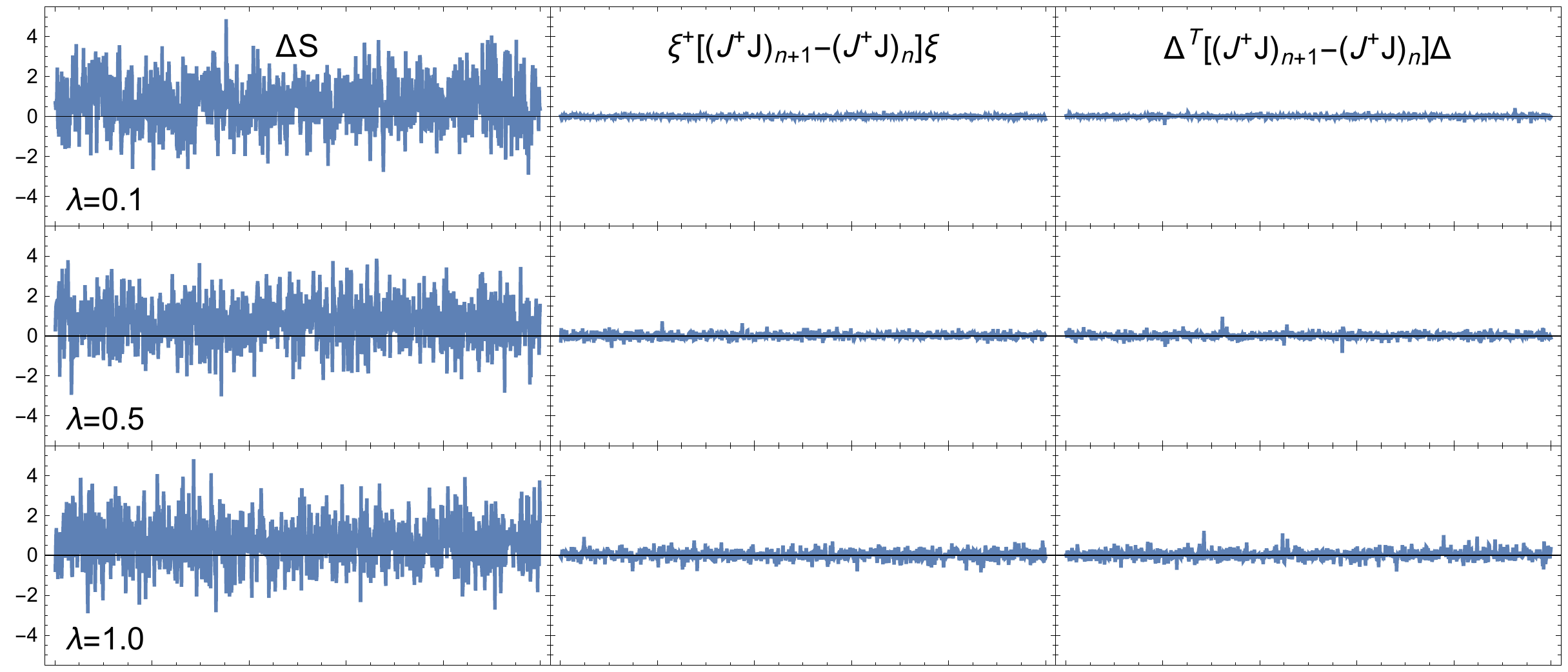} 
\caption{Time histories for a thermalized configurations using Grady algorithm.
The coupling increases from the upper row to the lower. The three columns
indicate the components that enter the acceptance rate in Eq.~\ref{eq:pacc}:
action change, $\xi$-estimator, and measure variation.}
\label{fig:fluctuations}
\end{figure}

The first issue we will address is the performance of the estimator
that appears in the acceptance rate in Eq.~\ref{eq:pacc}. 
We can find the acceptance rate by integrating Eq.~\ref{eq:pxi} over $\xi$; 
the result is a 
ratio of determinants $\det(J^\dagger J)_{n+1}/\det(J^\dagger J)_n$. One
potential problem would be a low acceptance rate or, equivalently, the need to use very small proposals  $\delta$.
In Fig.~\ref{fig:fluctuations} we show the relative
size of the three components that enter in the acceptance rate: the action change
$\Delta S$, the $\xi$-estimator $\xi^\dagger [ (J^\dagger J)_{n+1}-(J^\dagger J)_n]\xi$, and 
the measure change $\Delta^T [(J^\dagger J)_{n+1}-(J^\dagger J)_n]\Delta$. We
see that the action change dominates both the $\xi$-estimator and the measure
change, for all values of the coupling. We conclude that there is very
little loss associated with using the $\xi$-estimator rather than the
determinant ratio in the acceptance rate.

The observable that we will focus on in this section is the correlator
\beq
C(t,p) = \av{\phi(t, p) \phi(0, p)^\dagger}_\beta
\quad\text{with}\quad
\phi(t, p) \equiv \frac1{N_x} \sum_{n=0}^{N_x-1} e^{i p n} \phi_{t,n} \,.
\eeq
Note that the field $\phi$ is real thus $\phi(p)^\dagger = \phi(-p)$.
We use periodic fields in the spatial direction so the momentum is quantized
in units of $2\pi/N_x$.
To compute the correlator $C(t,p)$ we compute the averages on the forward
time leg of the Schwinger-Keldysh contour. To boost our statistics, we
can also use the fields on the backward leg of the contour. Note that the correlator
\beq
C_b(t,p) = \av{\phi(0,p)^\dagger \phi(t,p)}_\beta \,,
\eeq 
can be evaluated using the fields on the backward leg due to the reverse
time ordering of the operator product. We assume here that $t\geq 0$.
This correlator is related to $C$ via the relation 
$C(t,p)=\overline{C_b(t,-p)}$. Thus the observable we use is the following
\beq
C(t,p) = \frac1{2(N_t+1-t)} \sum_{t'=0}^{N_t-t} 
\left[\av{\phi(t+t',p) \phi(t',-p)} + 
\overline{\av{\phi(t'+T_b, p) \phi(t+t'+T_b, -p)}} \right] 
\text{ for }t=0,\ldots,N_t\,,
\eeq
where $T_b=N_t+1+N_\tau/2$ is the beginning of the backward leg in our
discretization. The formula above exploits the time translation invariance
along the forward/backward legs. Note that this symmetry is exact only in
the continuum: our discretization breaks this slightly due to the corner
effects.

\begin{figure}[t]
\includegraphics[width=0.457\textwidth]{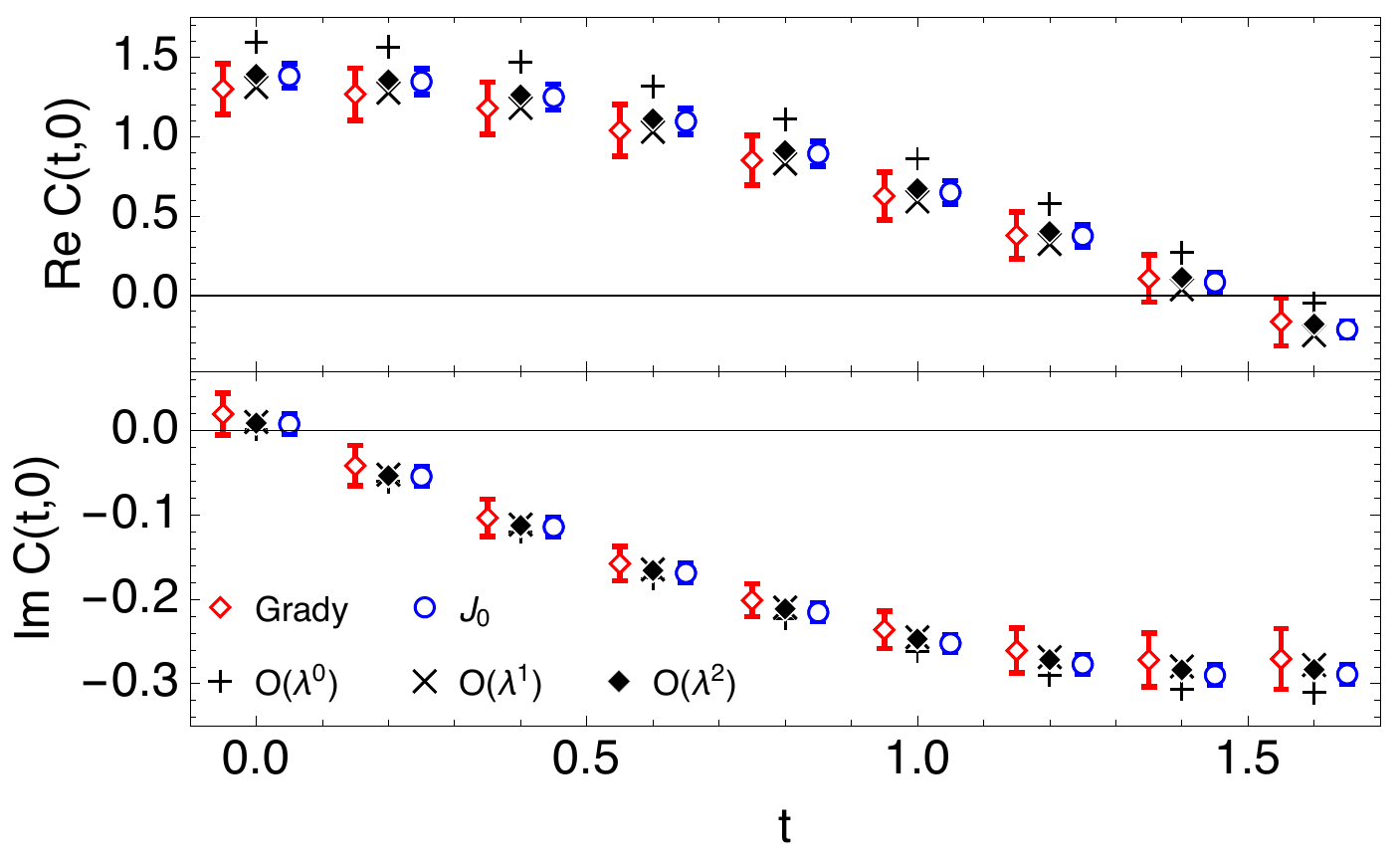} 
\kern0.3cm\includegraphics[width=0.443\textwidth]{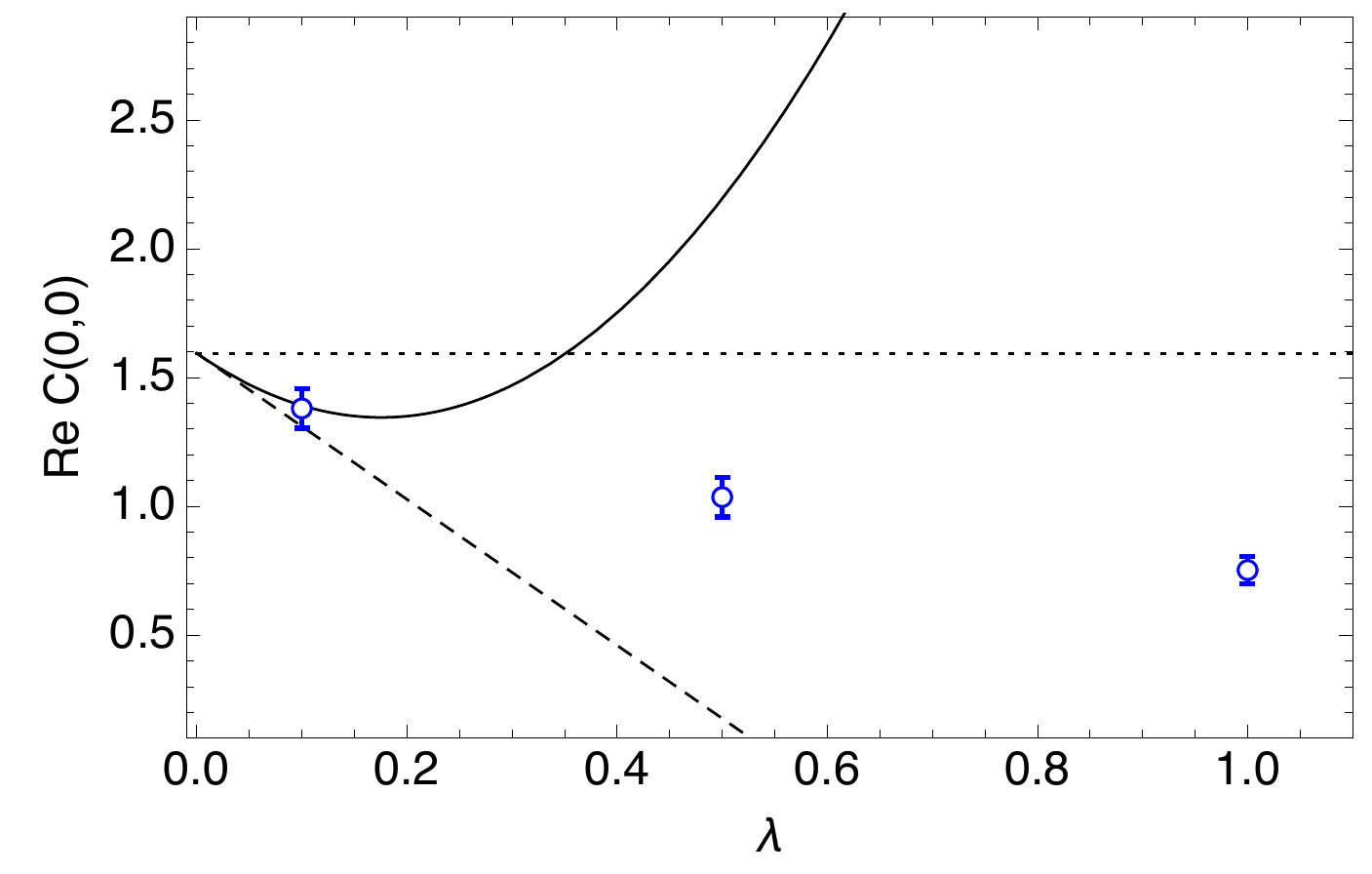}
\caption{Left: real and imaginary part of the correlator for $\lambda=0.1$
for momentum $p=0$,
as produced with the Grady and $J_0$ algorithms, compared to the perturbative
calculation. The simulation points are offset horizontally for clarity.
Right: the results for zero distance correlator as a function of the coupling.
The blue points are the results of the ``$J_0$" algorithm  and the curves correspond
to zeroth, first, and second order calculation.}
\label{fig:lam0p1}
\end{figure}

In the left panel of Fig.~\ref{fig:lam0p1} we compare the results of our 
simulations with the perturbative calculation
in the weak coupling ($\lambda=0.1$)  and $p=0$ case. We see the the results
from our two algorithms agree with each other---the error bars are smaller 
for $J_0$ because the number of statistics is significantly larger---and they agree 
very well with the perturbative results. Note that for this coupling
the perturbation theory seems to work well: the effect of higher order 
terms gets smaller. 
For larger values of $\lambda$, however, the perturbative results should become unreliable. We quantify the convergence of perturbation theory 
on the right panel 
of Fig.~\ref{fig:lam0p1} where we plot the value of $C(t=0,p=0)$ as a function of $\lambda$ at orders $\lambda^0, \lambda^1$ and $\lambda^2$. We see
that the series becomes unreliable well before we get to $\lambda=0.5$,
with the second order term having larger magnitude than the first order
correction, and the series diverging sharply away from our simulation points at $\lambda=0.5$ and 
$1.0$. We conclude that $\lambda=0.5$ and $1.0$ simulations are in the
strong coupling region.

\begin{figure}[t]
\includegraphics[width=0.45\textwidth]{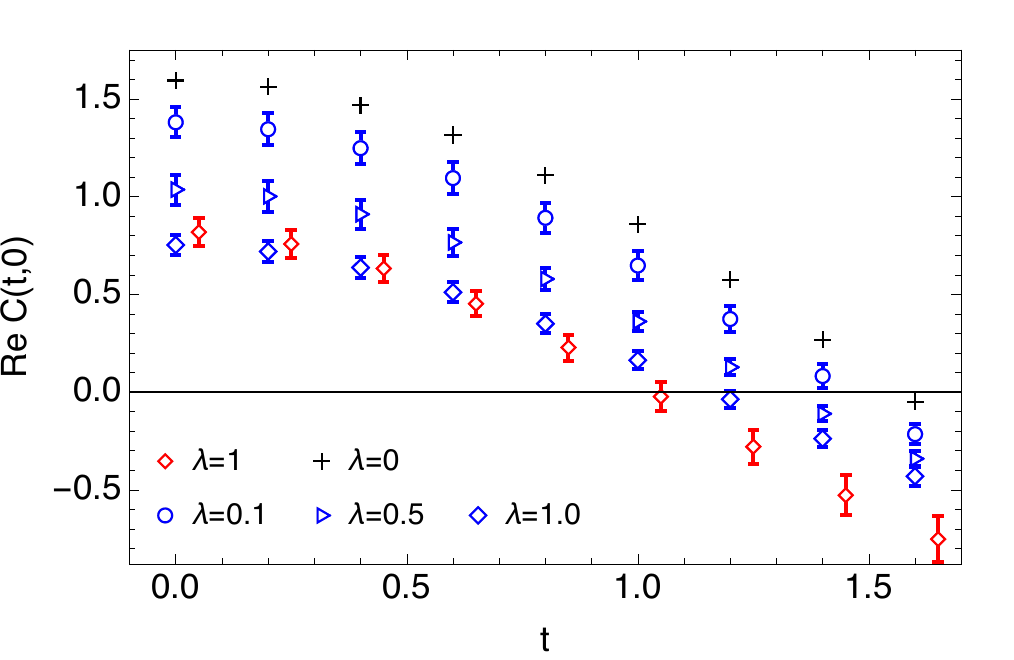} 
\kern0.3cm\includegraphics[width=0.45\textwidth]{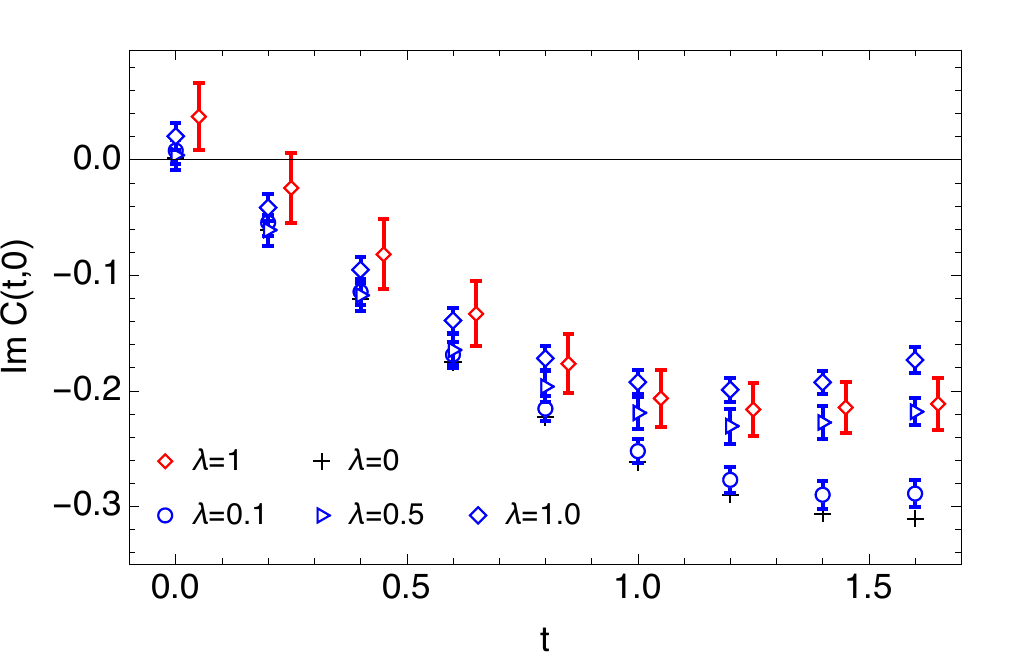}
\caption{Real and imaginary part of the correlator for momentum $p=0$ for
all $\lambda$ values used in this study.
The blue points are the results produced with $J_0$ 
algorithm and the red points correspond to the Grady algorithm.
For clarity the Grady results are displaced horizontally.
The black crosses are the result of exact calculations at $\lambda=0$.
}
\label{fig:alllam}
\end{figure}

In Fig.~\ref{fig:alllam} we plot the results for the real part of the
correlator for all values of $\lambda$ used in this study. Since the
statistics for the simulations using the Grady algorithm are smaller,
the error bars are larger, so we only display the results for the
largest coupling $\lambda=1.0$ (the ones for $\lambda=0.1$ are included
in Fig~\ref{fig:lam0p1}.) We see that the correlator is systematically
drifting away from its $\lambda=0$ value as we increase the value of
the coupling. We have run statistical tests and determined that the results of the Grady and $J_0$ algorithm are statistically
compatible. The apparent discrepancy in Fig.~\ref{fig:alllam} is not 
statistically significant since the data-points for different time
separations are strongly correlated.

\begin{figure}[t]
\includegraphics[width=0.457\textwidth]{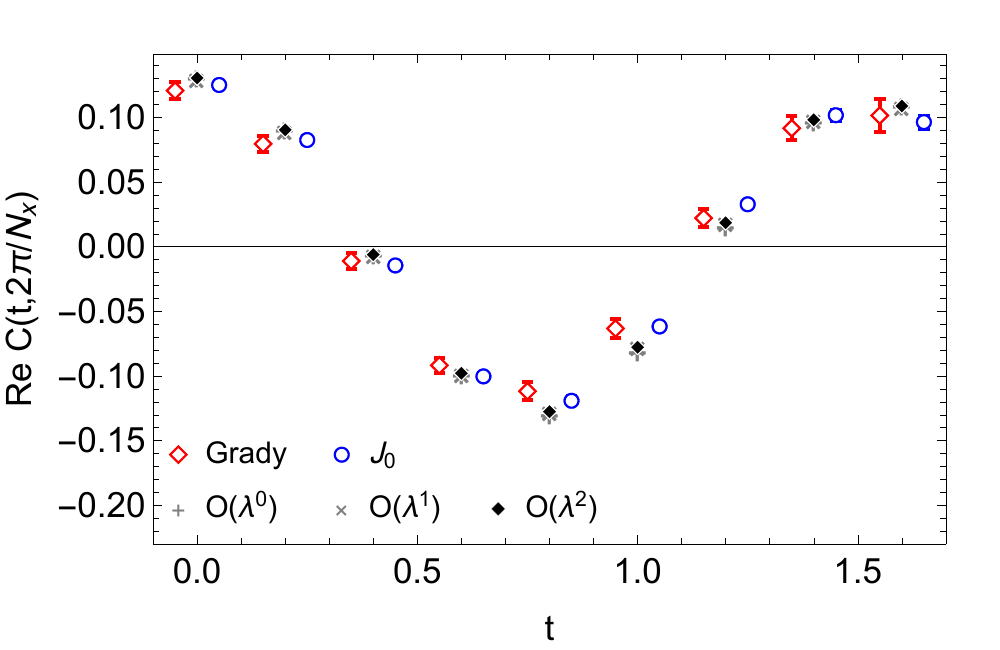} 
\kern0.3cm\includegraphics[width=0.443\textwidth]{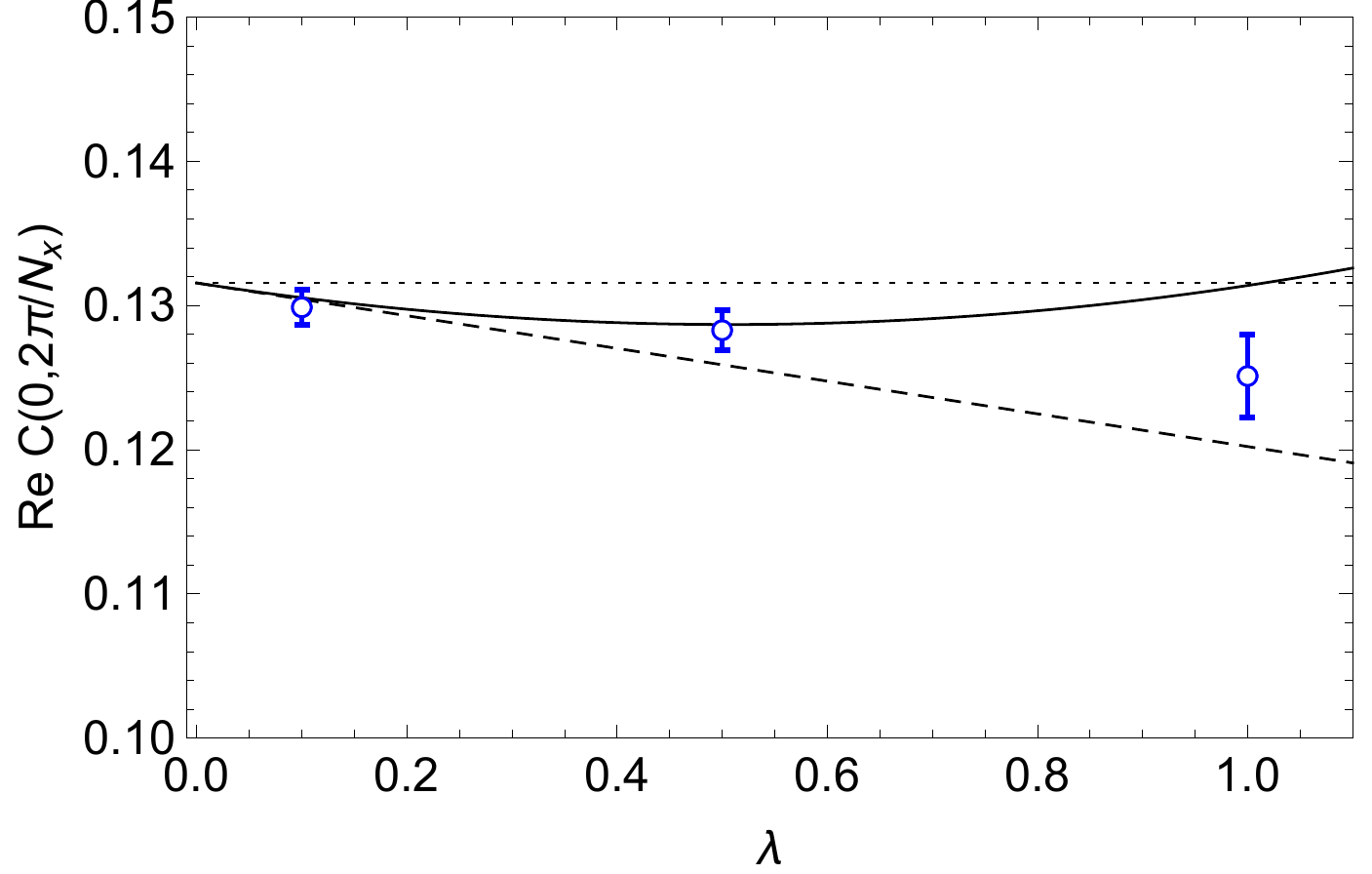}
\caption{Left: real part of the correlator for $\lambda=1.0$
for momentum $p=2\pi/N_x$,
as produced with the Grady and $J_0$ algorithms, compared to the perturbative
calculation. The simulation points are offset horizontally for clarity.
Right: the results for zero distance correlator as a function of the coupling.
The blue points are the results of $J_0$ simulation and the curves correspond
to zeroth, first, and second order calculation.}
\label{fig:mom1}
\end{figure}

In Fig.~\ref{fig:mom1} we show the correlator for the smallest non-zero
momentum, $p=2\pi/N_x$. We see that these correlators tend to 
have smaller statistical errors (this is true for higher momenta too),
and they change only slightly as we increase the strength of the
coupling. In the right panel of Fig.~\ref{fig:mom1} we show the
expectations from perturbation theory for this correlator: we see
that the first and second order effects are much smaller than for
the zero-momentum case, in agreement with the results of our simulations.

Before we conclude, we note that the performance of the ``$J_0$" algorithm
has been better than we anticipated. This was mainly due to the fact that
for the parameters studied in this paper the fluctuations of the Jacobian
were small. Thus the reweighting from the probability distribution $P_0$
sampled by the $J_0$ algorithm to the distribution $P(x)$ sampled by the Grady 
algorithm was very successful. To quantify this we measure the statistical
power for each of the ensembles generated with $J_0$ algorithm
\beq
\text{stat power} = \frac1{N_\text{cfg}}
\frac{\left( \sum_i |\det J(\phi_i)| \right)^2}
{\sum_i |\det J(\phi_i)|^2} \,,
\eeq 
where $N_\text{cfg}$ is the number of configurations in the ensemble, and
$\phi_i$ are the configurations. This quantity is equal to one when all
configurations contribute equally, which happens when $|\det J(\phi_i)|$ does
not fluctuate at all, and in the worst case it is $1/N_\text{cfg}$ when one
configurations has a dominant contribution to the reweigthed ensemble.
The statistical power of each ensemble is listed in Table~\ref{tab:simulations}.
We see that as expected the statistical power decreases as  we increase
$\lambda$, but the its value is still $0.68$ even on the ensemble 
with $\lambda=1.0$, so that we can easily reweight.


\section{Discussion and conclusions}
\label{sec:conclusion}

We have explored two different algorithms to compute the path integrals arising in the Schwinger-Keldysh formalism. They are both based on deforming the contour of integration from real variables to a submanifold of the complexified field space. They both bypass the most difficult and costly part of the computation, namely, the calculation of the jacobian of the parametrization of the deformed submanifold by real variables. They also lead to more efficient, isotropic Monte Carlo proposals.

The first algorithm has general applicability and can be seen as an adaptation of the Grady algorithm previously proposed to deal with the fermion determinant. The second (``$J_0$"), uses a free field approximation of the jacobian (with the difference between the correct and the free field jacobian reweighted during measurements). This last algorithm breaks down at strong enough coupling but we observed that it performs efficiently well before the point where perturbation theory is no longer valid.

The algorithms were applied to the computation of real time thermal correlators in the $1+1$ dimensional $\phi^4$ scalar theory. The two methods agreed with each other and with perturbation theory results at small enough values of the coupling. The ``$J_0$" algorithm is very efficient and its success at even relatively large values of the coupling is somewhat surprising. These calculations are, to our knowledge, the first reliable Monte Carlo real time calculations in a field theory.

The algorithms developed in this paper paves the way for larger scale calculations with finer lattices and/or larger number of spatial dimensions. The extension of the maximum time (bound in the present paper by $4\beta$) is a little more subtle and we uncover some evidence of trapping of the Monte Carlo chain in local minima of the effective action. In this case, the methods advocated in \cite{Fukuma:2017fjq,Alexandru:2017oyw} should be useful and should be incorporated.

\acknowledgments 

A.A. is supported in part by the National Science Foundation CAREER 
grant PHY-1151648 and by U.S. Department of Energy grant DE-FG02-95ER40907. 
A.A. gratefully acknowledges the hospitality of the Physics Departments 
at the Universities of Maryland and Kentucky, and the Albert Einstein Center 
at the University of Bern where part of this work was carried out. 
P.F.B. and G.R.  are supported by U.S. Department of Energy under Contract No. DE-FG02-93ER-40762. G.B. is supported by U.S. Department of Energy under Contract No. DE-FG02-01ER41195.

\bibliography{realtime-MC,thimbology}

\end{document}